\documentclass[iop,numberedappendix]{emulateapj-rtx4}
\usepackage{graphicx,natbib,color,bm,url,times,slashbox}
\graphicspath{{./fig/}{./png/}}

\newcommand{\EQ}{\begin{equation}}
\newcommand{\EN}{\end{equation}}
\newcommand{\EQA}{\begin{eqnarray}}
\newcommand{\ENA}{\end{eqnarray}}

\newcommand{\Eq}[1]{Equation~(\ref{#1})}
\newcommand{\Eqs}[2]{Equations~(\ref{#1}) and~(\ref{#2})}

\newcommand{\Sec}[1]{Section~\ref{#1}}

\newcommand{\Fig}[1]{Figure~\ref{#1}}
\newcommand{\FFig}[1]{Figure~\ref{#1}}
\newcommand{\FFigp}[2]{Figure~\ref{#1}({#2})}
\newcommand{\Figp}[2]{Figure~\ref{#1}({#2})}
\newcommand{\Figsp}[3]{Figures~\ref{#1}({#2}) and ({#3})}

\newcommand{\Tab}[1]{Table~\ref{#1}}

\newcommand{\bra}[1]{\langle #1\rangle}

{}
{}
{}

{}
{}
{}
{}
{}
{}
{}
{}
{}
{}
{}
{}
{}
{}
{}
{}
{}

{}

{}

{}
{}
{}

\newcommand{\grav}{\bm{g}}
\newcommand{\BB}{\bm{B}}

\newcommand{\JJ}{\bm{J}}

\newcommand{\AAA}{\bm{A}}

\newcommand{\UU}{\bm{U}}

\newcommand{\ff}{\mbox{\boldmath $f$} {}}

\newcommand{\nab}{{\bm{\nabla}}}

\newcommand{\SSSS}{\mbox{\boldmath ${\sf S}$} {}}

\newcommand{\ii}{{\rm i}}

\newcommand{\DD}{{\rm D} {}}

\newcommand{\dd}{{\rm d} {}}

\def\degr{\hbox{$^\circ$}}

\def\cs{c_{\rm s}}

\def\kf{k_{\rm f}}

\def\urms{u_{\rm rms}}

\def\half{{\textstyle{1\over2}}}

\def\onethird{{\textstyle{1\over3}}}

\newcommand{\G}{\,{\rm G}}

\newcommand{\uG}{\,\mu{\rm G}}
\newcommand{\um}{\,\mu{\rm m}}

\newcommand{\millim}{\,{\rm mm}}
\newcommand{\cm}{\,{\rm cm}}

\newcommand{\nm}{\,{\rm nm}}
\newcommand{\m}{\,{\rm m}}

\newcommand{\Mm}{\,{\rm Mm}}

\newcommand{\pc}{\,{\rm pc}}

\newcommand{\AU}{\,{\rm AU}}

\newcommand{\yapj}[3]{ #1, {ApJ,} {#2}, #3}

\newcommand{\yapjl}[3]{ #1, {ApJ,} {#2}, #3}

\newcommand{\yana}[3]{ #1, {A\&A,} {#2}, #3}

\newcommand{\yjfm}[3]{ #1, {J.\ Fluid Mech.,} {#2}, #3}

\newcommand{\yjetp}[3]{ #1, {Sov.\ Phys.\ JETP,} {#2}, #3}

\newcommand{\ymn}[3]{ #1, {MNRAS,} {#2}, #3}

\newcommand{\ysph}[3]{ #1, {Solar Phys.,} {#2}, #3}

\newcommand{\yssr}[3]{ #1, {Spa.\ Sci.\ Rev.,} {#2}, #3}

\newcommand{\yjour}[4]{ #1, {#2}, {#3}, #4}

\begin{document}
\title{Compensating Faraday depolarization by magnetic helicity in the solar corona}
\author{Axel Brandenburg$^{1,2,3,4}$, Mohira B.\ Ashurova$^{1,2}$, \& Sarah Jabbari$^5$
}
\affil{
$^1$Laboratory for Atmospheric and Space Physics, University of Colorado, Boulder, CO 80303, USA\\
$^2$Department of Astrophysical and Planetary Sciences, University of Colorado, Boulder, CO 80303, USA\\
$^3$Nordita, KTH Royal Institute of Technology and Stockholm University, Roslagstullsbacken 23, SE-10691 Stockholm, Sweden\\
$^4$Department of Astronomy, AlbaNova University Center, Stockholm University, SE-10691 Stockholm, Sweden\\
$^5$School of Mathematical Sciences and Monash Centre for Astrophysics,
Monash University, Clayton, VIC 3800, Australia
}
\submitted{Astrophys. J. 845, L15 (2017)}
\date{Received 2017 June 28; revised 2017 July 28; accepted 2017 August 4; published 2017 August 17}

\begin{abstract}
A turbulent dynamo in spherical geometry with an outer corona is simulated
to study the sign of magnetic helicity in the outer parts.
In agreement with earlier studies, the sign in the outer corona
is found to be opposite to that inside the dynamo.
Line-of-sight observations of polarized emission are synthesized to explore
the feasibility of using the local reduction of Faraday depolarization to
infer the sign of helicity of magnetic fields in the solar corona.
This approach was previously identified as an observational
diagnostic in the context of galactic magnetic fields.
Based on our simulations, we show that this method can be successful in
the solar context if sufficient statistics are gathered by using averages
over ring segments in the corona separately for the regions north and
south of the solar equator.
\end{abstract}

\keywords{
Sun: magnetic fields --- dynamo --- magnetohydrodynamics --- turbulence}
\email{brandenb@nordita.org}

\section{Introduction}

The solar magnetic field has an opposite twist in the two hemispheres.
This is seen, for example, in H$\alpha$ images of the Sun through
the orientation of sigmoidal structures of filaments in absorption.
These structures are {\sc S}-shaped in the south and {\sc N}-shaped
in the north, thus revealing a clear hemispheric dependence \citep{Mar03}.
A similar dependence is also seen in the twist of force-free magnetic
fields extrapolated from vector magnetograms around active regions
\citep{See90,PCM95}.
These indicate negative (positive) helicity in the northern (southern)
hemisphere.
The same hemispheric sign dependence was confirmed previously using
magnetic helicity spectra that were computed from solar surface vector
magnetograms \citep{ZBS16,BPS17}.

Magnetic helicity spectra have also been computed from time series of
the magnetic field vector measured on board the {\em Ulysses} spacecraft as it
flew at high northern and southern heliographic latitudes \citep{BSBG11}.
However, the signs of magnetic helicity turned out to have the opposite
sign of what is measured at the solar surface.
This was rather surprising, although it could be understood as a
consequence of a subdominance of generating effects (e.g., the
$\alpha$ effect in dynamo theory) compared with dissipating effects
(turbulent magnetic diffusion) in the solar wind.
These two effects tend to affect the sign of magnetic helicity in
opposite ways.
In the convection zone, the $\alpha$ effect is dominant, but in the
solar wind it is expected to be subdominant.
This unusual sign reversal of magnetic helicity was then confirmed by
\cite{WBM11} using numerical simulations of a turbulent helical dynamo
driven in the two hemispheres of a spherical wedge with a quiescent
exterior.
The current helicity, a proxy of magnetic helicity at small scales, was
found to be positive (negative) in the northern (southern) hemisphere,
i.e., just the other way around than in the dynamo region.
They interpreted this in a slightly modified way by arguing that in the
northern (southern) hemisphere, the dynamo sheds negative (positive)
magnetic helicity through a turbulent diffusive helicity flux
\citep{HB11}.
Analogous to Fickian diffusion of temperature, a flux is carried by a
negative gradient, but here the magnetic helicity can have either sign.
Thus, the negative magnetic helicity of the dynamo in the northern
(southern) hemisphere is carried by a positive (negative) magnetic helicity
gradient, driving it toward and arguably through zero.
This would explain the opposite sign of magnetic helicity some distance
above the solar surface.
If this idea is indeed applicable to the Sun, it would be
important to find out the distance above the solar surface, where the
change of sign occurs.
Could it be detected, for example, with Parker Solar Probe as it approaches
the Sun down to $0.04\AU$, or could the sign reversal be measured
within the solar corona ($0.01\AU$),
or perhaps even right at the solar surface?

Attempts to determine coronal magnetic helicity through morphological
considerations and force-free extrapolations \citep{PVDD15,Valori16}
or by measuring helicity flux through the surface \citep{Kazachenko}
may be biased toward large scales.
An alternate technique 
could utilize the effect of Faraday rotation along the line of sight.
In the absence of magnetic helicity, a line-of-sight magnetic field leads
to Faraday rotation and thus the superposition of polarization vectors
with different orientations, which is called Faraday depolarization.
A helical magnetic field of suitable sign can have the opposite effect
and thus compensate Faraday depolarization and therefore increase the
polarized intensity \citep{Soko98,BS14,HF14}.
A helical field of opposite sign leads to a decrease in polarized intensity.
Specifically, a line-of-sight magnetic field
pointing toward (away from) the observer would decrease Faraday rotation,
and thus enhance polarized intensity of suitable wavelength, if the
magnetic field has positive (negative) magnetic intensity \citep{BS14}.
This result has been known in the galactic context, where the radiation
is due to synchrotron emission.
In the solar context, we have to rely on polarized radiation from
magnetic-dipole transitions that occur in the corona
at certain discrete wavelengths \citep{Jud98,CJ99}.
\cite{DGRTJ11} proposed the use of polarized emission to infer
the twisted nature of coronal magnetic fields through forward modeling
of the Stokes vector and comparing against measurements with
the Coronal Multichannel Polarimeter
(CoMP) telescope \citep{Tom08,DKB16,Gib_etal17}.
However, Faraday rotation was not invoked in their approach, which
would require longer wavelengths in the millimeter range, as will be
shown below.

For the Sun, using narrow bandwidth observations at $\lambda=6\cm$
radio wavelengths, \cite{ACD94} found an oscillatory variation of the
Stokes $Q$ and $U$ parameters with respect to small changes in $\lambda$.
However, those wavelengths are too long to
determine magnetic helicity.
Furthermore, we also need the line-of-sight magnetic field,
because it determines the Faraday depolarization.
This can be obtained by determining the rotation measure,
i.e., the derivative of the
polarization angle with respect to wavelength, giving the sign of the
toroidal magnetic field.
This is another standard concept used mainly in radio astronomy,
but it applies to other wavelengths as well.
The correlation between rotation measure and polarized intensity is
therefore a direct proxy of magnetic helicity and was first proposed
by \cite{VS10}.
We emphasize that with our technique the actual orientation
of the transverse component of the magnetic field is not important.
It is only the {\em change} of the orientation with increasing distance
from the observer that enters.
In particular, no background sources are invoked and only the radiation
from within the corona is used.
The purpose of this Letter is to discuss the feasibility of this technique
in the solar context and to apply it to a simple model such as that
of \cite{WBM11,WBM12}.

\section{Description of the method}
\label{DescriptionMethod}

The simplest example we can construct is that of a Beltrami field, which
\cite{BS14} wrote as $(B_x,B_y,B_z)=(B_\perp\sin kz,B_\perp\cos kz,0)$,
where the observer is in the negative $z$ direction.
Here, $k$ is the wavenumber of the magnetic field.
They expressed the component perpendicular to the direction of the
observer $\BB_\perp=(B_x,\,B_y)$ in a complex
form as ${\cal B}\equiv B_x+\ii B_y=r_B\exp(\ii\psi_B)$.
In the present arrangement, the observer is in the negative $y$ direction,
so we rotate $z\to y$, $B_y\to B_x$, and $B_x\to B_z$, so we have
\EQ
\BB=\pmatrix{B_x\cr B_y\cr B_z}=\pmatrix{
B_\perp\!\cos ky\cr B_{\|0}\cr B_\perp\!\sin ky},
\label{Beltrami}
\EN
with $\BB_\perp=(B_x,\,B_z)$ and
${\cal B}\equiv B_z+\ii B_x=r_B\exp(\ii\psi_B)$.
We have assumed here a constant line-of-sight magnetic field,
$\BB_\|=(0,B_{\|0},0)$.
The intrinsic linear polarization vector $(q,\,u)$ is then
\EQ
q+\ii u=p_0 \epsilon \exp(2\ii\psi_p),
\EN
where $\psi_P=\psi_B+\pi/2$ is the electric field
angle, $\epsilon(x,y,z)$ is the emissivity, and $p_0$
is the degree of polarization.
Integrating along the line of sight yields the observable polarization,
written here in complex form as
\EQ
P(x,z,\lambda^2)\equiv Q+\ii U=p_0\int_{-\infty}^\infty \epsilon\,
e^{2\ii(\psi_P+{\phi}\lambda^2)}\,\dd y,
\label{Pint}
\EN
where $\lambda$ is the wavelength,
\EQ
\phi(x,y,z)=-K\int_{-\infty}^y n_{\rm e}(x,y',z)\,B_\|(x,y',z)\,\dd y'
\label{phiz}
\EN
is the Faraday depth, with $n_{\rm e}$ being the electron density
and $K=0.81\m^{-2}\cm^3\uG^{-1}\pc^{-1}=2.6\times10^{-17}\G^{-1}$
being a constant \citep[e.g.,][]{ACD94}.
As in \cite{BS14}, we assume $\epsilon\propto B_\perp^\sigma$
and compare $\sigma=2$ and 0.
Furthermore, we normalize $P$ by the total intensity
$I=\int\epsilon\,\dd y$.
Of particular interest is the case when the polarized emission is maximum,
which is when the exponent in \Eq{Pint} vanishes.
\Eq{phiz} applies to nonuniform $n_{\rm e}$ and $B_\|$, but we
now discuss the case when $n_{\rm e}=n_{\rm e0}$ and $B_\|=B_{\|0}$
are constants.
A fully helical magnetic field of the form given by \Eq{Beltrami} makes
the exponent vanish if $\psi_B-\pi/2=-ky$, i.e., if
the wavenumber of the field in \Eq{Beltrami} obeys
\EQ
k=-K n_{\rm e0} B_{\|0} \lambda^2.
\EN
In that case, Faraday depolarization becomes minimal, i.e., we have
maximum polarization.
This is the essence of this technique.

To get an idea about the ranges in $\lambda$ and $n_{\rm e}$ that would
be needed to obtain cancelation for a magnetic field of wavenumber
$k=0.01\Mm^{-1}=1500\AU^{-1}$,
which corresponds to a length scale of $(2\pi/0.01)\Mm\approx600\Mm$,
we have listed plausible combinations of $n_{\rm e}$, $\lambda$, and $B_\|$
in \Tab{Tsum}.
This wavenumber lies on the lower end of values relevant to the solar
surface \citep{BPS17} and near the upper end of values in the solar
wind \citep{BSBG11}.
Thus, the far- to near-infrared wavelength range is optimal for
detecting helical magnetic fields.
On a scale of $60\Mm$, all wavelengths would be three times larger.
To discuss the feasibility of this method further, we determine
the line-of-sight integrated polarization using the magnetic field from
a simulation similar to that of \cite{WBM11}.

\begin{table}[t!]\caption{
Wavelength $\lambda$ of maximum polarized emission for fully helical magnetic
fields with $k=0.01\Mm^{-1}=1500\AU^{-1}$ for $n_{\rm e}$ $[\cm^{-3}]$
and $B_\|$ $[\G]$.
}\vspace{12pt}\centerline{\begin{tabular}{lccccccc}
\backslashbox{$B_\|\!\!$}{$\!\!n_{\rm e}$} & $10^6$ & $10^8$ &
$10^{10}$ & $10^{12}$ & $10^{14}$ \\
\hline
$0.01\G$ & $20\cm$    & $2\cm$     & $2\millim$ & $200\um$ & $20\um$  \\
$1\G$    & $2\cm$     & $2\millim$ & $200\um$   & $20\um$  & $2\um$   \\
$100\G$  & $2\millim$ & $200\um$   & $20\um$    & $2\um$   & $200\nm$ \\
\label{Tsum}\end{tabular}}\end{table}

\section{Numerical simulations}
\label{NumericalSimulations}

We solve the hydromagnetic equations for the magnetic vector potential
$\AAA$, the velocity $\UU$, and the logarithmic density
$\ln\rho$, using an isothermal equation of state
with constant sound speed $\cs$,
\EQ
{\partial\AAA\over\partial t}=\UU\times\BB-\eta\mu_0\JJ,
\EN
\EQ
{\DD\UU\over\DD t}=\grav+\ff-\cs^2\nab\ln\rho
+{1\over\rho}\left[\JJ\times\BB+\nab\cdot(2\nu\rho\SSSS)\right],
\EN
\EQ
{\DD\ln\rho\over\DD t}=-\nab\cdot\UU,
\EN
where $\BB=\nab\times\AAA$ is the magnetic field,
$\JJ=\nab\times\BB/\mu_0$ is the current density,
$\mu_0$ is the magnetic permeability, $\nu$ is the kinematic viscosity,
${\sf S}_{ij}=\half(U_{i,j}+U_{j,i})-\onethird\delta_{ij}\nab\cdot\UU$
is the rate-of-strain tensor, $\grav$ is the gravitational acceleration,
and $\ff$ is a forcing function; see below.

We consider a wedge-shaped computational domain in spherical coordinates
$(r,\vartheta,\varphi)$ with
\EQ
0.7\leq r/R\leq 2,\quad
15\degr\leq\vartheta\leq165\degr,\quad
-17.5\degr<\varphi<17.5\degr,
\EN
where $R$ is the solar radius.
The gravitational acceleration is $\grav=(-GM/r^2,0,0)$,
where $G$ is Newton's constant and $M$ is the solar mass.
We use $GM/(R\cs^2)=3$, which results in a density contrast of about 16
in the radial direction.
As in \cite{WBM11}, $\ff$ consists of plane waves
with typical wavenumber $\kf=3k_0$ and is nonvanishing only in the
``turbulence zone'' in $0.7\leq r/R\leq 1$.
Here, $k_0=2\pi/(0.3\,R)$ is the lowest radial wavenumber in this zone
of thickness $0.3\,R$.
The helicity of $\ff$ changes sign about the equator and is
negative (positive) in the northern (southern) hemisphere.
We use the {\sc Pencil Code}\footnote{\url{https://github.com/pencil-code}}
in spherical wedge geometry with $144\times288\times72$ mesh points
in the $r$, $\vartheta$, and $\varphi$ directions.

The magnetic field grows at first exponentially with time at a growth
rate $\gamma\approx0.073\,\tau^{-1}$, where $\tau=(\urms\kf)^{-1}$
is the turnover time in the dynamo region of our model and is about
$\tau=0.14\,R/\cs$.
The magnetic field develops a cycle with equatorward migration.
The period is about $2000\,\tau$, which is about 10 times longer than for
the smaller wedges of \cite{WBM11}, which spanned $\pm18\degr$ latitude.
Such migratory dynamos without differential rotation were discovered
by \cite{MTKB10}.
In contrast to earlier work \citep{WBM11,WBM12}, we have now extended
the latitude range to $\pm75\degr$.
Models with this latitudinal extent, but no corona, where also studied
by \cite{JBKMR15}, who investigated the spontaneous formation of spots at
the surface in the presence of dynamo action, but at much larger
stratification.

Our model is different from the standard scenario of a solar dynamo,
which involves differential rotation.
One reason for adopting an $\alpha^2$ dynamo is its simplicity, while
capturing essential features of a realistic turbulent dynamo:
scale separation, different signs of magnetic helicity at large and
small scales, and magnetic helicity fluxes out of the domain
and across the equator.
As a model for the Sun, such a dynamo is not implausible
\citep{KMCWB13,MS14}.
However, as we will see below, in our model the magnetic field is
strongest at high latitudes.
This could in principle be alleviated by adopting a
modified helicity profile, as done in \cite{JBKMR15}.
Such refinements, as well as the inclusion of differential rotation,
would be useful extensions for future work.

\section{Calculation of the line-of-sight magnetic field}

To perform line-of-sight integrations as in
\Eqs{Pint}{phiz}, we overlay a Cartesian mesh with coordinates $(x,y,z)$,
and look up at each Cartesian meshpoint the nearest magnetic field value
on the spherical mesh at position $(r,\vartheta,\varphi)$.
The components of $\BB=(B_r,B_\vartheta,B_\varphi)$ are then expressed
in terms of Cartesian components.
As in \Sec{DescriptionMethod},
the observer is assumed to be looking in the positive $y$ direction.
Thus, $B_\varphi>0$ implies positive $B_\|=B_y$ in the first or fourth
quadrants, which corresponds to negative Faraday depth; see \Eq{phiz}.

\begin{figure}[t!]\begin{center}
\includegraphics[width=\columnwidth]{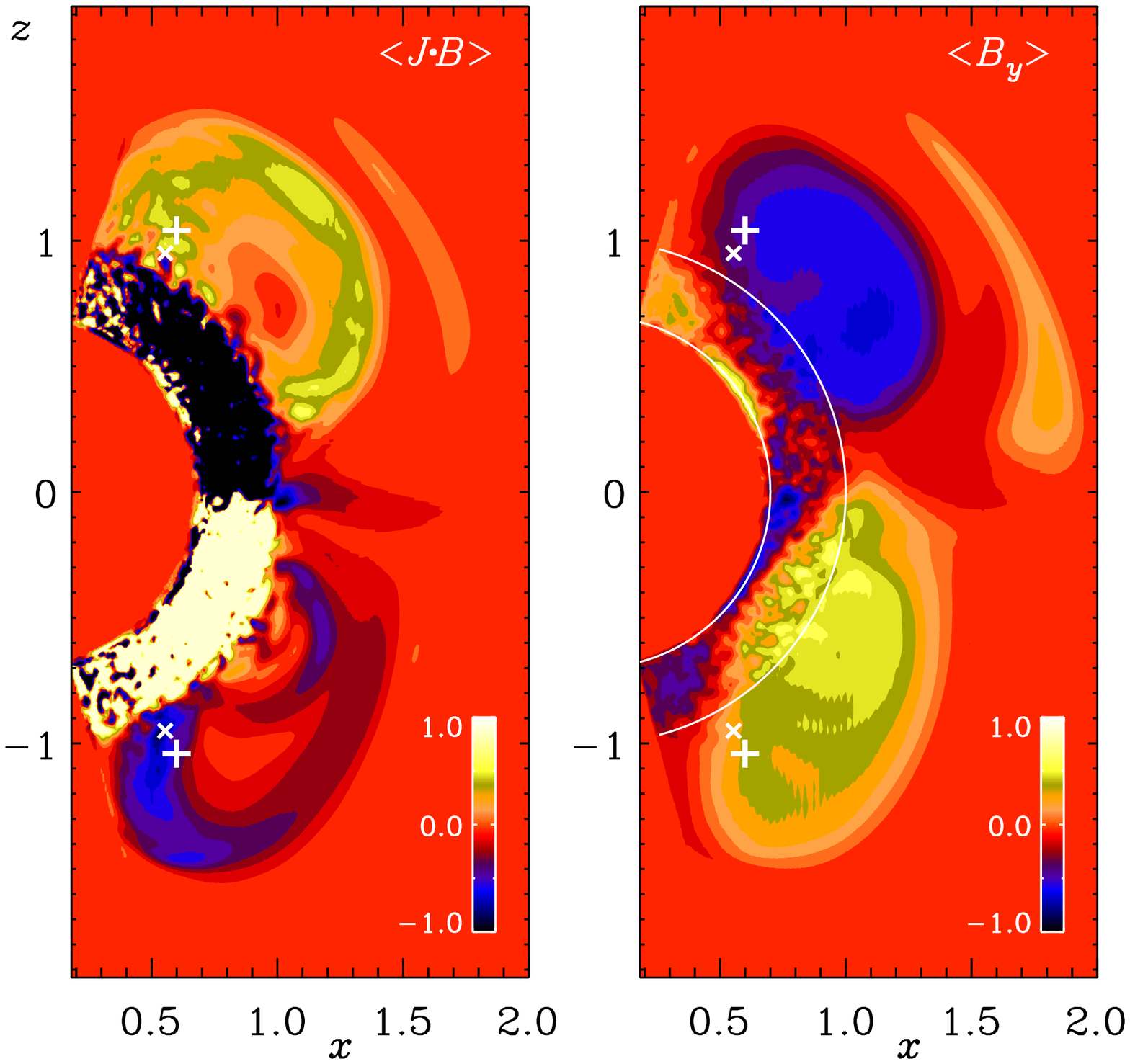}
\end{center}\caption[]{
$\bra{\JJ\cdot\BB}_y$ (left) and $\bra{B_y}_y$ (right) at $t/\tau=430$.
The crosses and plusses mark positions at $r/R=1.1$ and $1.2$ and
$90\degr-\vartheta=\pm60\degr$ latitude, for which the $\lambda^2$
dependence is studied in \Fig{ppfft2_spectra}.
}\label{ppjbm_bym}\end{figure}

In \Fig{ppjbm_bym}, we plot the current helicity $\bra{\JJ\cdot\BB}_y$
and mean toroidal field $\bra{B_y}_y$,
where $\bra{\cdot}_y$ denotes averaging along the line of sight.
In $r<R$, $\bra{\JJ\cdot\BB}_y$
is negative (positive) in the northern (southern)
hemisphere, but it changes sign for $r>R$ and becomes positive (negative)
in the northern (southern) hemisphere.
Furthermore, $\bra{B_y}_y$ is negative in the first quadrant (northern
hemisphere), so the Faraday depth is positive; see \Eq{phiz}.

\begin{figure}[t!]\begin{center}
\includegraphics[width=\columnwidth]{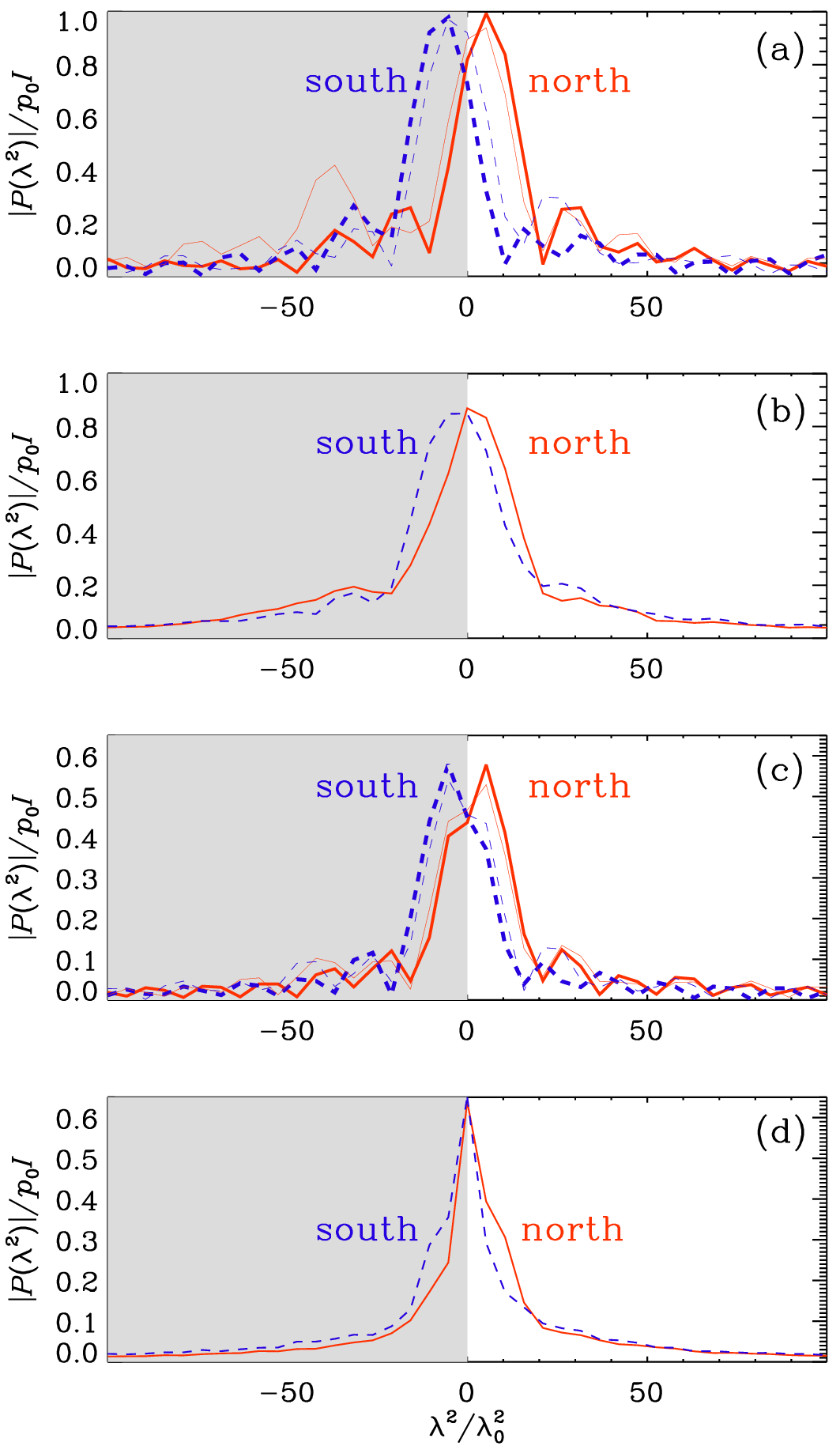}
\end{center}\caption[]{
(a) $|P(\lambda^2)|$ at four reference points indicated in the two panels
of \Fig{ppjbm_bym} in the northern (red) and southern (blue) hemispheres
at $t/\tau=430$.
Thick (thin) lines refer to $r/R=1.2$ ($1.1$) at $\pm60\degr$ latitude.
(b) $|P(\lambda^2)|$ averaged over radial shells,
$r/R=1.10$--$1.15$.
Shaded areas denote $\lambda^2<0$, which are unphysical.
((c) and (d)) Same as (a) and (b), but for $\sigma=0$.
}\label{ppfft2_spectra}\end{figure}

\FFig{ppjbm_bym} shows that in the northern hemisphere, the coronal
magnetic field has positive $\bra{\JJ\cdot\BB}_y$.
This is consistent with the results of \cite{WBM11} and, since current
helicity is a proxy of small-scale magnetic helicity, it is also consistent
with the results for the solar wind \citep{BSBG11}.
Let us now ask whether this result can also be inferred from the polarized
intensity computed from our models using \Eq{Pint}.
We begin by plotting $|P(\lambda^2)|$ at points where the field is strongest.
As alluded to at the end of \Sec{NumericalSimulations}, this is in our model
at high latitudes, so we choose four reference points at $\pm60\degr$ latitude 
at $r/R=1.1$ and $1.2$ indicated in the two panels of \Fig{ppjbm_bym}.
The result is shown in \Figp{ppfft2_spectra}{a}, where we have normalized
$|P|$ by the total intensity $I$ at the same point, and $\lambda^2$ is
normalized by
\EQ
\lambda_0^2\equiv(K n_{\rm e0} B_{\|0} R)^{-1},
\EN
which implies that $\lambda^2/\lambda_0^2=kR$.
In this case, the values of $\lambda$ given in \Tab{Tsum} are somewhat
smaller: $0.75\cm$ instead of $2\cm$, for example.

We see from \Figp{ppfft2_spectra}{a} that, in the northern hemisphere,
the polarized intensity has a maximum at a positive value of
$\lambda^2/\lambda_0^2$.
This is consistent with our expectation that for positive Faraday depth,
i.e., negative $B_\varphi$, polarized intensity should be maximum
for positive values of $\lambda^2$ if the magnetic helicity is positive
\citep{BS14}.
In the southern hemisphere, \Eq{Pint} shows that the polarized intensity
has a maximum at negative values of $\lambda^2$, which is of course
unphysical and unobservable.
However, even in that case, the integral in \Eq{Pint} can still be evaluated.
In fact, it is well known that this integral is just the usual Fourier
integral provided the integration is performed over $\phi$ instead of $y$
\citep{Bur66,BB05}.
If $B_\varphi$ were positive (e.g., half a Hale cycle later),
one should see more polarized intensity in the south instead.

\begin{figure}[t!]\begin{center}
\includegraphics[width=\columnwidth]{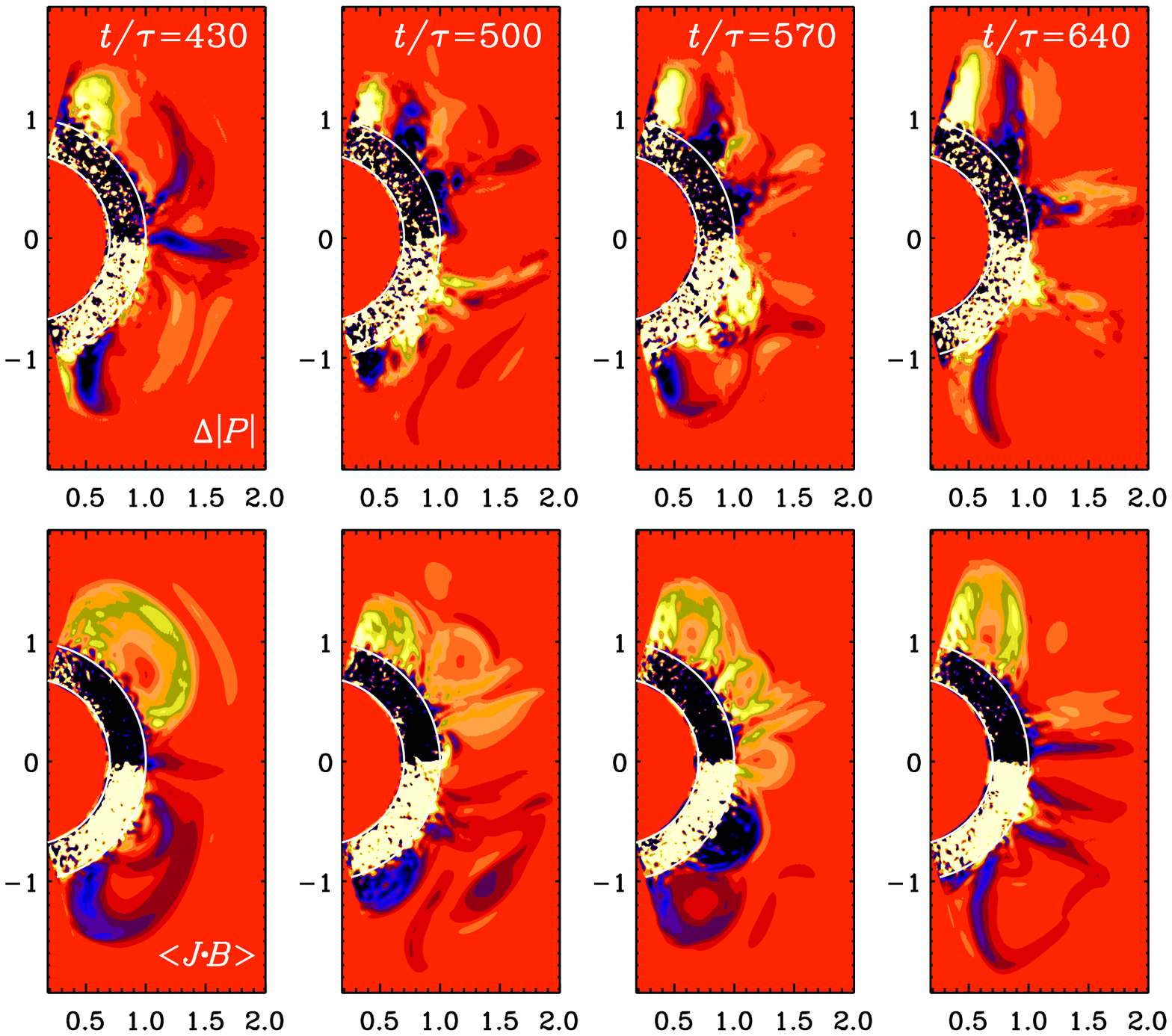}
\end{center}\caption[]{
$\Delta|P|$ (upper row) and $\bra{\JJ\cdot\BB}$
in the $xz$ plane of the observer at four times in the interval
$t/\tau=430$--$640$.
The color table is the same as in \Fig{ppjbm_bym}.
}\label{ppfft3_pi_times}\end{figure}

\FFigp{ppfft2_spectra}{a} shows that the maximum of $|P|/I$ is at
$\lambda^2/\lambda_0^2\approx5$, i.e., $\lambda/\lambda_0\approx2.2$.
For the Sun, at $r/R=1.1$, we expect $n_{\rm e}=10^{-8}\cm^{-3}$.
Using $B_{\|0}=1\G$, as an example, we have $\lambda_0\approx2\millim$, so
$\lambda\approx4\millim$, which is at the limit of ALMA.
In the outer parts, $n_{\rm e}$ would be lower, so $\lambda$ would
be larger still.
The results for $\sigma=0$ are similar to those for $\sigma=1$;
see \Figsp{ppfft2_spectra}{c}{d}.

Looking only at one position in the corona may not be enough to get a
reliable result about the coronal magnetic helicity.
In fact, as we will see further below, exceptions to the correspondence
between polarized intensity and current helicity are not uncommon.
Therefore, a more robust method is to use hemispheric ring averages,
$\bra{|P|}_{\rm(N/S)}$, which are averages of $|P(r,\theta)|$ over an
interval $r_1<r<r_2$ and $0<\theta<\pi/2$ for the north (N) and
$\pi/2<\theta<\pi$ for the south (S).
The result is shown in \Figp{ppfft2_spectra}{b} for a ring with $r_1/R=1.10$
and $r_2/R=1.15$.
The difference in polarized intensity for north and south
is now no longer so striking, but
it may well be good enough if sufficient statistics are gathered.

Incidentally, \Figp{ppfft2_spectra}{a} also shows oscillations in the wings
at larger values of $\lambda^2$ with $\Delta\lambda^2\approx20\lambda_0^2$.
This is a consequence of the finiteness of nonvanishing contributions to the
integral in \Eq{Pint} for a finite slab \citep{Bur66}.
Such oscillations have indeed been observed by \cite{ACD94} using radio
observations of the solar corona at small bandwidth at $6\cm$ wavelength.
In our simulation, this corresponds to a slab of width
$L=2\pi R/20\approx0.3\,R$, which agrees with our domain size
along the line of sight.

To demonstrate the relationship between helicity and polarized intensity
more thoroughly, we now consider an artificially constructed quantity
\EQ
\Delta|P|\equiv\bra{|P|}_+ - \bra{|P|}_-,
\EN
where the $\bra{|P|}_\pm$ denote the averages of $|P|$ over the
intervals $0<\lambda^2/\lambda_0^2<60$
and $-60<\lambda^2/\lambda_0^2<0$, respectively
Again, the negative $\lambda^2$ interval is of course unobservable
in reality, but computing it from our models
allows us to see more clearly the degree of
correspondence with the $\bra{\JJ\cdot\BB}$ maps.
In \Fig{ppfft3_pi_times} we show $\Delta|P|$ for four times separated
by $70\,\tau$ around the times considered above.
The visualizations of $\Delta|P|$ are found to be a reasonable proxy of
$\bra{\JJ\cdot\BB}$ inside the turbulence zone ($r<R$), but in the corona
$\Delta|P|$ is no longer a good proxy -- at least not at all times.
This, again, highlights the need for using averages to obtain
reliable results.

\section{Conclusions}

Our results have confirmed that there is a correspondence between
polarized intensity and magnetic or current helicity.
This idea was originally applied to galaxies, but it should also work
for the Sun using polarized emission from within the corona some distance
above the solar surface.
The most appropriate
wavelengths lie in the millimeter range, which has only now become
accessible through ALMA.

Using studies of polarized intensity to constrain the solar dynamo and
magnetic helicity in the corona may shed light on the nature of the
dynamo mechanism, which is likely to involve an $\alpha$ effect as
a result of cyclonic convection, as anticipated already by \cite{Par55}.
Such a dynamo produces helical magnetic fields through
an inverse cascade of magnetic helicity \citep{PFL76}.
However, unlike kinetic helicity, magnetic helicity is conserved and both
positive and negative signs tend to be produced at the same time,
but at different length scales.
Different signs of magnetic helicity are also present
in the solar wind at large and small scales.
\cite{BSBG11} associated the helicity at the largest scales with that of
the Parker spiral \citep{Par58}, which is negative in the north \citep{BEM87}.
At smaller scales, the sign of magnetic helicity in the solar wind agrees
with that at large scales in the dynamo interior.
Our new simulations suggest that the apparent sign reversal may occur close
to the solar surface; see the lower panel of \Fig{ppfft3_pi_times}.
This raises our hopes that further guidance for our understanding
of this effect can come from observations.

In the present work, we have examined the possibility of using the
compensating effect of a helical magnetic field on Faraday rotation.
This idea has not yet received much attention in solar physics, except
for early work of the 1990s that showed the essence of Faraday rotation
at radio wavelengths \citep{ACD94}.
These authors considered observations on the solar disk above active
regions, but solar limb observations appear plausible too.
It is essential to use a broad range of wavelengths from infrared
to millimeter wavelengths.
However, the actual location of this helicity reversal should be treated
with care.
It is therefore essential to inspect a suitable range of data using
not only ALMA and CoMP observations, but also in situ observations using,
for example Parker Solar Probe to inspect statistical properties of the
field at close range.

\acknowledgements
We thank the referee for constructive comments and Gabriel Dima,
David Elmore, Phil Judge, and Padma Yanamandra-Fisher
for useful discussions.
This work has been supported in part by
the NSF Astronomy and Astrophysics Grants Program (grant 1615100),
the Research Council of Norway under the FRINATEK (grant 231444),
and the Swedish Research Council (grant 621-2011-5076).
We acknowledge the allocation of computing resources provided by the
Swedish National Allocations Committee at the Center for Parallel
Computers at the Royal Institute of Technology in Stockholm.
This work utilized the Janus supercomputer, which is supported by the
National Science Foundation (award number CNS-0821794), the University
of Colorado Boulder, the University of Colorado Denver, and the National
Center for Atmospheric Research. The Janus supercomputer is operated by
the University of Colorado Boulder.
The input files as well as some of the output files of the simulation
are available under
{http://lcd-www.colorado.edu/~axbr9098/projects/spherical-surface}.


\vfill\bigskip\noindent\tiny\begin{verbatim}
$Header: /var/cvs/brandenb/tex/mohira/spherical_surface/paper.tex,v 1.44 2017/08/17 16:08:31 brandenb Exp $
\end{verbatim}

\end{document}